
\input harvmac
\noblackbox
%
%

%
-

\def\Title#1#2{\ifx\answ\bigans \nopagenumbers
\abstractfont\hsize=\hstitle\rightline{#1}%
\vskip .5in\centerline{\titlefont #2}\abstractfont\vskip
.5in\pageno=0
\else \rightline{#1}
\vskip .8in\centerline{\titlefont #2}
\vskip .5in\pageno=1\fi}
\ifx\answ\bigans
 
scaled\magstep3
 
scaled\magstep3
 
scaled\magstep3
 
scaled\magstep3
 
scaled\magstep3
\else
 
scaled\magstep3
 
scaled\magstep3
 
scaled\magstep3
 
scaled\magstep3
 
scaled\magstep3
 
 \font\absi=cmmi10 scaled\magstep1
\font\absis=cmmi7 scaled\magstep1 \font\absiss=cmmi5 scaled\magstep1
\font\abssy=cmsy10 scaled\magstep1 \font\abssys=cmsy7 scaled\magstep1
\font\abssyss=cmsy5 scaled\magstep1 
scaled\magstep1
\skewchar\absi='177 \skewchar\absis='177 \skewchar\absiss='177
\skewchar\abssy='60 \skewchar\abssys='60 \skewchar\abssyss='60
\fi
%
%

\def\hmu{\hat \mu}
\def\hnu{\hat \nu}
\def\hone{\hat 1}
\def\htwo{\hat 2}
\def\hthree{\hat 3}
\def\hfour{\hat 4}
\def\hf{{1\over2}}

\def\ajou#1&#2(#3){\ \sl#1\bf#2\rm(19#3)}

\def\quart{{1 \over 4}}
\def\Tr{{\rm Tr}}

\def\frac#1#2{{#1 \over #2}}

\def\real{I\negthinspace R}
\def\zed{ Z\hskip -3mm Z }
%
\lref\nohair{Ref. for no hair Theorems}
\lref\bw{D. R. Brill and
J. A. Wheeler, ``Interactions of Neutrinos and Gravitational
Fields,'' \ajou
Rev. Mod. Phys. &29 (57) 465.}
\lref\eweinberg{E. Weinberg, \ajou Phys. Rev. &D24 (81) 2669.}
\lref\growhair{S. Coleman, J. Preskill, and F. Wilczek, ``Growing
Hair on
Black Holes,'' \ajou Phys. Rev. Lett. &67 (91) 1975.}
\lref\qhair{S. Coleman, J. Preskill, and F. Wilczek, ``Quantum Hair
on
Black Holes'' IASSNS-HEP-91/64, CALT-68-1764, HUTP-92/A003, (1991)
and
references therein.}
\lref\jr{R. Jackiw and P. Rossi, ``Zero Modes of the Vortex-Fermion
System,''
\ajou Nucl. Phys. &B190 (81) 681.}
\lref\witten{E. Witten, ``Superconducting Strings,'' \ajou Nucl.
Phys. &B249
(85) 557.}
\lref\dgt{F. Dowker, R. Gregory, and J. Traschen, ``Euclidean Black
Hole
Vortices,''
FERMILAB-Pub-91/142-A, EFI-91-70 (1991).}
\lref\bghhs{M. J. Bowick et. al.}
\lref\PSSTW{J. Preskill, ``Quantum hair'', CALT-68-1671 (1990);
J. Preskill, P. Schwarz, A. Shapere, S. Trivedi and
F. Wilczek, ``Limitations on the Statistical Description of Black
Holes,''
IAS preprint IASSSNS-HEP-91/34.}
\lref\wittenbh{E.Witten, \ajou Phys. Rev. D &44 (91) 314.}

\lref\bghhs{M.Bowick, S.Giddings, J.Harvey, G.Horowitz and
A.Strominger,
\ajou Phys. Rev. Lett. &61  (88) 2823.}

\lref\kw{L.Krauss and F.Wilczek, \ajou Phys. Rev. Lett. &62 (89)
1221.}

\Title{\vbox{\baselineskip12pt
\hbox{EFI-92-15} }}
{\bf QUANTUM FERMION HAIR}
\baselineskip=12pt
\bigskip
\centerline{Ruth Gregory and Jeffrey A. Harvey}
\bigskip
\centerline{\it Enrico Fermi Institute, University of Chicago}
\centerline{\it 5640 Ellis Avenue, Chicago IL 60637}
\bigskip
\centerline{\bf Abstract}

It is shown that the Dirac operator in the background of a magnetic
Reissner-Nordstr\"om
black hole and a Euclidean
vortex possesses normalizable zero modes in theories containing
superconducting cosmic strings.
One consequence of these zero modes  is the presence of a fermion
condensate around magnetically charged black holes which violates
global quantum numbers.

\bigskip
\medskip
\Date{4/92}
\eject
\newsec{Introduction}
Classically black holes are among the most perfect objects in the
universe,
being traditionally characterized according to the classical no hair
theorems
by only
their mass, angular momentum, and electric and magnetic charges.
It has been realized recently that the structure of black holes
is possibly much richer. Quantum mechanically black holes can carry
a variety of hair which can be detected by means of Aharonov-Bohm
interference
effects \bghhs. In \bghhs\ axion hair which can be detected
through scattering of axion strings or superstrings off black holes
was discussed. In \kw\ it was argued that discrete gauge hair
could also be carried by black holes and a number
of consequences of such hair were explored in \growhair, \qhair,
\dgt.

One of the striking conclusions of \growhair\ was the fact that
Euclidean vortex solutions in a black hole background
give rise to a  non-perturbative, exponentially
decaying electric field outside the horizon of a Schwarzschild
black hole. Thus in principle the effects of quantum hair can be
detected not only by interference effects, but directly through
measurement of the electric field.
The vortex solutions which give rise to these
effects are mathematically very similar to  cosmic string solutions
which arise in spontaneously broken $U(1)$ gauge theories. It has
been known for some time that fermion fields in the presence of
such cosmic strings can have normalizable zero modes \jr. In some
models
these zero modes lead to superconducting cosmic strings \witten.
It is natural to wonder whether these zero modes can also play
a role in the structure of black holes. In this paper we answer this
in
the affirmative for magnetically charged Reissner-Nordstr\"om black
holes
by showing that operators constructed out of fermion fields
acquire non-zero expectation values outside the horizon.

The outline of the paper is as follows. The second section contains a
review
of the relevant results regarding cosmic string superconductivity. In
the third section we discuss euclidean vortices in black hole
backgrounds
and extend previous results to the case of Reissner-Nordstr\"om
solutions.
In the fourth section we combine
and extend these results to demonstrate the existence of fermion zero
modes in the combined black hole-vortex configuration. In the fifth
section we discuss some physical effects of these zero modes, in
particular
we show that magnetic black holes develop a fermion condensate
outside
the horizon which violates global quantum numbers.

\newsec{Cosmic Strings and Fermion Zero Modes}

The model of fermion zero modes of interest
to us was introduced by Jackiw and Rossi in \jr\ and generalized
by Witten in \witten. This theory (in flat Minkowski space) has
a $U(1)_R$ gauge symmetry with gauge field strength $G_{\mu \nu} =
\partial_\mu
B_\nu - \partial_\nu B_\mu$, a
fermion field $\psi$ with charge $r_\psi=-1/2$, and a charge one
scalar
field $\phi$. The lagrangian is given by
\eqn\jrlag{{\cal L} = -{1 \over 4} G^{\mu \nu} G_{\mu \nu} +
  i \bar \psi  \gamma^\mu D_\mu \psi + \half i \lambda
\phi  \psi^T C \psi - \half i  \lambda \phi^*
 \bar \psi C  \bar \psi^T + D_\mu \phi^*
D^\mu \phi -V(\phi),}
where $V(\phi)$ is a potential for $\phi$ with a minimum at
$|\langle \phi \rangle |=v$,  the covariant derivative is given by
$D_\mu = (\partial_\mu + i e_R R B_\mu)$ with $R$ the $U(1)_R$
charge,
$C$ is the
charge conjugation matrix,
we take $\lambda$ to be real and positive,
and we may for simplicity choose $\psi$ to have definite chirality.

It is then convenient to use a chiral basis for the gamma matrices
with
\eqn\gammas{\gamma^5 = \left( \matrix{-1 & 0 \cr
                                       0 & 1 \cr} \right) \qquad
            \gamma^\mu = \left( \matrix{ 0 & \sigma^\mu \cr
                                         \bar \sigma^\mu & 0 \cr}
\right)
\qquad
            C = \left( \matrix{ -i\sigma^2 & 0 \cr
                                  0 & i \sigma^2 \cr} \right) }
and to set
\eqn\psichoice{\psi = \left( \matrix{ \psi_L \cr 0 \cr} \right) }
where $\sigma^\mu = (1,\sigma^i)$, $\bar \sigma^\mu = (1,-\sigma^i)$
in terms of the usual Pauli matrices.

A static vortex configuration running along the $z-$axis
with winding number $k=1$ is given in polar
coordinates $(\rho,\theta)$ by
\eqn\vorconfig{\eqalign{ \phi(\rho) &= v X_0(\rho) e^{i \theta} \cr
                          B_\theta &= {1 \over e_R} (P_0(\rho) -1)
\cr}}
with $X_0(\rho)\to 0$ and $P_0(\rho)\to 1$ as $\rho \rightarrow 0$,
while
as $\rho \rightarrow \infty$
$P_0$ approaches zero and $X_0$ approaches unity.

The $\psi$ equation of motion following from  \jrlag\ is
\eqn\diraceqn{i \bar \sigma^ \mu D_\mu \psi_L +
\lambda \phi^* \sigma^2 \psi_L^* =0.}
A static zero-energy solution to the $\psi$ equation of motion in the
background \vorconfig\ will thus obey the equation
\eqn\zmodeeq{\vec \sigma \cdot ( i \vec \nabla - r_\psi e_R \vec B )
\psi_L -
             \lambda \phi^* \sigma^2 \psi_L^* = 0.}
Going to polar coordinates and writing the two-component spinor
$\psi_L$ as
\eqn\psidecomp{ \psi_L = \left( \matrix{
e^{r_\psi \int_0^\rho d \rho'(P_0-1)/ \rho'} \psi_U \cr
e^{- r_\psi \int_0^\rho d \rho' (P_0-1)/ \rho' } \psi_D \cr}
\right) }
yields the decoupled equations
\eqn\decoup{\eqalign{ e^{-i \theta} (i \partial_\rho \psi_D +
                      {1 \over \rho} \partial_\theta \psi_D)
             + i \lambda \phi^* \psi_D^* =& 0 \cr
                      e^{i \theta} (i \partial_\rho \psi_U -
                      {1 \over \rho} \partial_\theta \psi_U)
             - i \lambda  \phi^* \psi_U^* =& 0. \cr }}

For a single vortex configuration given by \vorconfig, \decoup\ has
a single normalizable solution $\psi_L^0$ given by $\psi_U^0 =0$,
\eqn\fzmode{ \psi_D^0 = e^{-\lambda v\int_0^\rho X_0(\rho') d \rho' }
}
which obeys $i \gamma_1 \gamma_2 \psi_L^0 = -\psi_L^0$.
In \jr\ Jackiw and Rossi analyze the spectrum of the Dirac operator
in a more general $k$-vortex configuration and show that it includes
precisely $k$ normalizable zero modes in accordance with the general
index theorem of \eweinberg.

Returning to the full equation of motion \diraceqn\ we can look for
solutions of the form
\eqn\zmode{\psi_L = \eta(z,t) \psi_L^0(\rho,\theta).}
Since $\gamma_5 \psi_L = i \gamma^0 \gamma^1 \gamma^2 \gamma^3 \psi_L
= -\psi_L$ and $i \gamma^1 \gamma^2 \psi_L^0 = -\psi_L^0$,
\diraceqn\ implies that
$\eta(z,t)= \eta(z+t)$ is a two-dimensional Majorana-Weyl spinor
which
represents fermions localized on the vortex which propagate down the
vortex at the speed of light.

In order to obtain charged zero-modes on the vortex it is necessary
to
double the degrees of freedom so as to obtain complex zero modes.
Thus
following \witten\ we can generalize the lagrangian \jrlag\ to
include
an additional left-handed fermion field $\chi$ and an additional
unbroken $U(1)_Q$ gauge symmetry with field strength $F_{\mu \nu} =
\partial_\mu A_\nu -
\partial_\nu A_\mu$. The lagrangian is
\eqn\witlag{\eqalign{{\cal L} = & - \quart F_{\mu \nu} F^{\mu \nu}
- \quart G_{\mu \nu} G^{\mu \nu} + i \bar \psi \gamma^\mu D_\mu \psi
+ i \bar \chi \gamma^\mu D_\mu \chi + D_\mu \phi^* D^\mu \phi -
V(\phi) \cr
& + i \lambda \phi \psi^T C \chi
- i \lambda \phi^*  \bar \psi C  \bar \chi^T .}}
The $U(1)_R$ symmetry is spontaneously broken by the expectation
value
of $\phi$ with $Q=0, R=+1$ and again gives rise to vortex solutions
as given by
\vorconfig. The Yukawa couplings are consistent with
the choice of quantum number assignments $Q=q, R=r$ for $\psi$ and
$Q=-q, R=-r-1$ for $\chi$,
although note that ${\rm Tr} QR= 2q(r+1/2)$ so that $Q$ is orthogonal
to
$R$ only for $r=-1/2$.\foot{We could of course work with
orthogonal $U(1)$ factors
to start with in which case the unbroken $U(1)$ would in general be a
linear combination of the original $U(1)$ factors.}

The $\psi$ and $\chi$ equations of motion are
\eqn\paireqn{\eqalign{ i \bar \sigma^\mu D_\mu \psi_L + \lambda
\phi^* \sigma^2
                        \chi_L^* &= 0, \cr
			i \bar \sigma^\mu D_\mu \chi_L + \lambda
\phi^* \sigma^2
			\psi_L^* &=0.}}
Repeating the previous analysis we now find a normalizable zero mode
in a $k=1$
vortex background given by
\eqn\czmode{ \chi_L = e^{-i \alpha} \psi_L^0 x_-(\rho) \;\;\; ,
\;\;\; \psi_L = e^{i \alpha} \psi_L^0 x_+(\rho) \; ,}
with $\alpha$ an arbitrary phase, and where $x_\pm \in$ \real, and
satisfy
\eqn\awkward{ x_\pm' \pm r_- {(P_0-1)\over \rho} x_\pm \pm \lambda v
X_0(\rho)
(x_- - x_+) = 0.}
$r_-= (r_\psi - r_\chi)/2= r +1/2$ represents the overlap
of U(1)$_Q$ with U(1)$_R$.
We have not been able to solve \awkward\ analytically, other than as
a perturbation series in $r_-$, but it is easy
to see that there exists a  solution with asymptotic behavior
\eqn\asympt{ x_+ \simeq 1 - { r_- (1-r_-) \over 2 \lambda v \rho} ,
\qquad
              x_- \simeq 1 +{ r_- (1+r_-) \over 2 \lambda v \rho},}
as $\rho\to\infty$, which makes the zero modes \czmode\ normalizable.

Thus we see that the previous real Majorana-Weyl zero mode has been
extended
to a complex Weyl zero mode which carries charge under the unbroken
$U(1)_Q$.

As emphasized in \witten, as it stands both the original
four-dimensional
theory
and the two-dimensional low-energy effective theory of the zero modes
is
anomalous. This may be addressed by adding an additional multiplet
of left-handed spinors $\hat \psi$, $\hat \chi$ with quantum
numbers $Q=\hat q$, $R=\hat r$ and $Q=-\hat q$, $R=-\hat r +1$
respectively which acquire mass from coupling to ${\phi}^*$ rather
than $\phi$:
\eqn\hatcoup{\hat {\cal L} = i \gamma \phi^* \hat
\psi^T C \hat \chi + {\rm h.c.}}
In the general situation with $M$ pairs $(\psi_i, \chi_i)$ and $N$
pairs
$(\hat \psi_j, \hat \chi_j )$ the conditions for anomaly cancellation
are
\eqn\anomcan{\eqalign{\Tr R^3 &=0 =\sum_{i=1}^M (-3r_i^2 -3r_i -1) +
                                   \sum_{j=1}^N (3 \hat r_j^2 -3 \hat
r_j +1)
				       ;\cr
	      \Tr Q^2 R &=0 =-\sum_{i=1}^M q_i^2 + \sum_{j=1}^N \hat
q_j^2
		                      ; \cr
	      \Tr Q R^2 &=0 =-\sum_{i=1}^M q_i(2 r_i+1) +\sum_{j=1}^N
\hat
                                      q_j (2 \hat r_j -1). \cr}}

The simplest anomaly free model has $M=N=1$ with $\hat q = q$ and
$\hat r = r+1$. An
embedding of this structure in a grand unified theory is given
by the $O(10)$ vortex model discussed in \witten.
In this model a vortex is formed in the breaking of $O(10)$ to
the standard model
and the dynamics require that the ordinary electroweak Higgs
doublet have a phase change of $2 \pi$ in going around the vortex.
The fields and their quantum numbers
are then (for a single generation) $\psi_1=e_L$, $\chi_1 = e_L^c$
with
$q_1=-1$, $r_1=1$; $\psi_2 = d_L$,
$\chi_2 = d_L^c$ with $q_2=-1/3$, $r_2=-1/3$ and $\hat \psi_1 = u_L$,
$\hat \chi_1=u_L^c$ with $\hat q_1=2/3$, $\hat r_1=-1/3$.
The zero modes in this
model consist of left-moving down quarks and anti-quarks, left-moving
electrons
and positrons, and right-moving up quarks and anti-quarks.

\newsec{Euclidean Black Hole Vortices}

In the following section we will discuss fermion zero modes
in a background consisting of a Euclidean vortex in a
Reissner-Nordstr\"om background. In this section we discuss the
basic features and existence of such a vortex solution.
In order to search for a vortex-Reissner-Nordstr\"om solution
we would ideally couple the Euclidean matter action
to gravity, employing suitable boundary
conditions which indicate the presence of a vortex.
In particular the vortex fields must have an
asymptotic winding number
\eqn\wind{
\eqalign{
\phi &\to  v e^{2\pi ik\tau /\beta} \cr
B &\to {-2\pi k\over  e_R \beta} d\tau \cr } }
as $r\to\infty$, and be regular as $r\to r_+$, the horizon.
The formalism for dealing with such a system
was developed in \dgt, although there it was used for a Schwarzschild
black hole. The Reissner-Nordstr\"om black hole represents an
increase
in calculational complexity, although many of the qualitative
features
of the vortex solution remain. Let us first review the salient points
of \dgt\ before proceeding to the
Reissner-Nordstr\"om vortex.

In \dgt\ it was noted that existence of solutions to the non-linear
coupled vortex-gravity equations is in general a very difficult
problem.
What enabled a tractable solution to be found was the observation
that
the situation was very similar to that of a self-gravitating cosmic
string,
and \dgt\ proved that the black hole vortex satisfied similar
criteria
to the cosmic string. Specifically, the metric was shown to be
asymptotically Schwarzschild,
{\it provided} the following conditions were satisfied:

{\it (i)} the system was weakly gravitating, and

{\it (ii)} the energy momentum decayed at least as fast as
${\hat r}^{-5}$ outside the core,

where
\eqn\rhat{
\hat r = \int_{r_+}^r {dr'\over A}
}
is the proper distance from the horizon.
This argument, coupled with a rewriting of the equations as
a perturbative expansion in $\epsilon$, the gravitational strength
of the vortex, and a numerical
background vortex solution allowed a strong plausibility
argument for existence of a fully coupled
Einstein-Abelian Higgs vortex solution. The back-reaction of
the vortex on the geometry was easily extracted, and was
shown to take the form of cutting a slice out of the
euclidean black hole cigar geometry - a gravitational effect
analogous to the self-gravitating cosmic string.

For a Reissner Nordstr\"om vortex, we expect that if we find a
background solution of a similar form to that in Schwarzschild,
then the rest of the argument will go through in an analogous
fashion. Since we are concerned with the physics of the vortex itself
here, rather than its gravitational effect, we restrict ourselves
to proving existence of a background solution. Questions of
gravitational backreaction and thermodynamics will be discussed
elsewhere.

In order to search for a vortex solution we make the usual
decomposition (with $B=\beta/2 \pi$)
\eqn\david{
\eqalign{
\phi &=   v X(r) e^{ik\tau/B} \cr
B_\mu &= {1\over Be} ( P(r) - k)\partial_\mu \tau = {1\over Be}
(P_\mu - k\partial_\mu \tau) \; , \cr
} }
which yields the equations of motion
\eqn\xandp{
\eqalign{
{1\over r^2} ( r^2 P_{,r})_{,r} &= {\lambda  v^2 \over \nu} {X^2 P
\over A^2} \cr
{1\over r^2} (r^2 A^2 X_{,r})_{,r} &= {P^2 X \over A^2 B^2} +
{\lambda  v^2\over 2} X(X^2 -1)\; , \cr
} }
where $A^2=g_{00}$, and $\nu = \lambda /2e^2$ is the Bogomoln'yi
parameter
representing the relative strengths of the Higgs and gauge
interactions. $\nu=1$ is often called the supersymmetric
limit.

The asymptotic behaviour of the background solutions to
\xandp\ is
\eqn\asymp{
\eqalign{
&X \propto (r-r_+)^{|k|/2} \hbox{\hskip 5mm} P = k -
O(r-r_+) \hbox{\hskip 5mm \rm as} \; r\to r_+ \cr
&1-X \propto r^{-1-a} e^{-\sqrt{\lambda} v r}
\hbox{\hskip 5mm} P \propto r^{-1-a/\sqrt{\nu}}
e^{-\sqrt{\lambda}  v r/\sqrt{\nu} }
\hbox{\hskip 5mm \rm as}
\; r\to\infty  \cr }
}
where $a=\sqrt{\lambda} v r_+(1+r_+/B)$.
In order to verify that these correspond to the asymptotic
behaviours of a full solution, we must integrate \xandp\ with
$A^2 = (r-r_+)(r-r_-)/r^2$. Setting
${\tilde r} = \sqrt{\lambda} v (r-r_+)$, $R=\sqrt{\lambda}v$
and $\Delta = 2Rr_+/B$ gives
\eqn\eqtosolv{
\eqalign{
{1\over ({\tilde r} +R)^2}
\left [ ({\tilde r} +R)^2 P' \right ] ' &=
{X^2P\over \nu} {( {\tilde r} +R)^2 \over {\tilde r}({\tilde
r} + \Delta) } \cr
{1\over ({\tilde r} + R)^2}
\left [ {\tilde r} ({\tilde r} +\Delta) X' \right ] ' &=
{P^2X\over 4R^4} {( {\tilde r} +R)^2 \Delta^2 \over {\tilde
r}({\tilde r} + \Delta) }
+ {\half} X (X^2-1). \cr
}
}
Note that $\Delta\to0$ gives extremal Reissner-Nordstr\"om,
and $R\to\infty$ gives the thin string limit. We have integrated
the equations numerically using a relaxation technique, and
the results show that the bounded solutions at the
horizon do indeed integrate out to the exponentially decaying
solutions at infinity. Figure 1a shows a plot of
$P$ with $k,\nu=1$, and $R=1$ for varying $\Delta$, and
Figure 1b a plot of $X$ for the same values of the
parameters. Note the
bunching up of the solutions near to the horizon as we lower
$\Delta$. This represents the divergence of the proper
distance to the horizon, which from \rhat\ is
\eqn\horn{
\eqalign{
{\hat r} &= \{ \sqrt{r-r_+} \sqrt{r-r_-} + (r_+ +
r_-) \log \left [ {\sqrt{r-r_+} + \sqrt{r-r_-} \over
\sqrt{r_+-r_-} } \right ] \} \cr
&\sim r_+\log ( {\tilde r} /\Delta) \;\;\;\;{\rm as} \;\;\;\;
\Delta\to0. }
}

As a final remark, note that if we replace $r$ by $\hat \rho=
\sqrt{\lambda}v \hat r$, \xandp\ becomes
\eqn\nova{
\eqalign{
{d\over d\hat\rho} \left ( {1\over \hat\rho} {dP\over d\hat\rho}
\right )
&= {1\over\nu} {X^2P\over \hat\rho} + O(\Delta/R^3) \cr
{d\over d\hat\rho} \left ( {\hat\rho} {dX\over d\hat\rho} \right )
&= {P^2 X \over\hat\rho} + \half {\hat\rho} X (X^2-1) +
O(\Delta/R^3).\cr
}
}
Thus either in the thin string or near extremal limit, the vortex
becomes
a Nielsen-Olesen vortex, and we can write
\eqn\xpform{
\eqalign{
X &= X_0 (\hat r) \cr
P &= P_0 (\hat r) \cr
}
}
where $X_0$, $P_0$ are the functions appearing in \vorconfig.

\newsec{Euclidean Fermion Zero Modes}

In this section we find fermion zero modes in a combined vortex
Reissner-Nordstr\"om background of the type discussed in the previous
section. In order to do so we must first discuss the euclidean
version
of the action \witlag\ in this background. In the following section
we will discuss the physical consequences of these zero modes.

First let us discuss \witlag.
In order to euclideanise we must set $t\to i\tau$, and
$g_{\mu \nu}\to-g_{\mu \nu}$
(since we were working with signature $+---$). We must then make an
overall
sign change to the Lagrangian in order to comply with the convention
that the
action for matter fields
\eqn\posact{ S_M = \int \sqrt{g} {\cal L}_M d^4x >0.}
In this case the Einstein equations take the form
\eqn\einstein{ G_{\mu \nu}=8\pi GT_{\mu \nu} = {16\pi G \over
\sqrt{g}}
{\delta ({\cal L}_M\sqrt{g})\over \delta g^{\mu \nu}} .}
Clearly the bosonic part of \witlag\ becomes
\eqn\bos{ {\cal L}_B = \quart F_{\mu \nu}^2 + \quart G_{\mu \nu}^2 +
|D_\mu \phi|^2 + V(\phi). }
For the fermionic part, note that it is only the spatial part of the
metric
that reverses sign, hence
\eqn\egam{ \gamma^\tau_E = \gamma^t_L \;\;\; , \;\;\;
\gamma ^i_E = i\gamma^i_L}
and therefore
\eqn\ferm{ {\cal L}_F = - \bar \psi \gamma^\mu D_\mu \psi
- \bar\chi \gamma^\mu D_\mu \chi - i \lambda ( \phi \psi^T C \chi
- \phi^*  \bar \psi C  \bar \chi^T).}

As is well known,
in euclidean space $\psi$ and $\bar \psi$ are no longer related
by complex conjugation and must be treated as independent fields.
`Lorentz'
invariance
in fact requires that $\bar \psi$ have opposite chirality to $\psi$.
Therefore in what follows we are implicitly taking $\psi,\chi$ to be
left-handed, and $\bar\psi,\bar\chi$ right-handed.

We thus wish to look for left-handed $\psi$ and right-handed
$\bar\chi$
solutions of the equations
\eqn\euczm{
\eqalign{ \gamma^\mu D_\mu \psi &= i\lambda\phi^* C \bar\chi^T \cr
\gamma^\mu D_\mu (\bar\chi^T)^* &= i\lambda\phi^* C \psi^* \cr } }
plus the analogous equations involving (right-handed) $\bar \psi$ and
(left-handed) $\chi$.

Our conventions are as follows. We will use coordinates
$x^1 = r$, $x^2 = \theta$, $x^3 = \phi$
and $x^4 = \tau$ to describe the Reissner-Nordstr\"om metric with
$(\theta,\phi)$ being angular coordinates on $S^2$, $r$ the radial
coordinate,
and $\tau$ the euclidean time coordinate. We will use hatted
indices for tangent space indices.
We start with a set of euclidean gamma matrices $\gamma^{\hmu}$,
$\hmu=1 \cdots 4$
obeying $\{\gamma^{\hmu} , \gamma^{\hnu} \} = 2 \delta^{\hmu \hnu}$.
An
explicit
chiral basis from \egam\ is given by
\eqn\eucgamma{\gamma^5 = \left( \matrix{-1 & 0 \cr
                                       0 & 1 \cr} \right) \qquad
            \gamma^{\hmu} = \left( \matrix{ 0 & \sigma^{\hmu} \cr
                              \bar \sigma^{\hmu} & 0 \cr} \right)
\qquad
            C = \left( \matrix{ -i\sigma^2 & 0 \cr
                                  0 & i \sigma^2 \cr} \right)  }
with $\sigma^{\hmu} = (i \sigma^i, 1)$ and
$\bar \sigma^{\hmu} = (-i \sigma^i,1)$. The curved space gamma
matrices
are related to these by $\gamma^\mu = {e^\mu}_{\hmu} \gamma^{\hmu}$
with ${e^\mu}_{\hmu}$
the vierbein.  The vierbein is then given by  ${e^\mu}_{\hmu}
= {\rm diag} (A(r),1/r,1/r \sin \theta, 1/A(r))$ with
\eqn\ar{ A^2(r) = (1 - {2 M \over r} + {4 \pi g_Q^2 \over r^2} )}
and $g_Q$ the magnetic charge.

The covariant derivative appearing in \euczm\ is given by
\eqn\covderiv{ D_\mu = \partial_\mu - \Gamma_\mu
                       +i R e_R B_\mu +i Q e_Q A_\mu }
with $\Gamma_\mu = -\hf {\omega_\mu}^{\hmu \hnu} \Sigma_{\hmu \hnu}$
in terms
of the spin connection ${\omega_\mu}^{\hmu \hnu}$ and
the euclidean Lorentz generators $\Sigma^{\hmu \hnu} = \quart
[\gamma^{\hmu} , \gamma^{\hnu}]$. $B_\mu$ is the broken
$U(1)_R$ gauge field with coupling constant $e_R$ and $A_\mu$ is the
unbroken
$U(1)_Q$ gauge field with coupling constant $e_Q$.

The background fields are given by the euclidean vortex configuration
\eqn\eucvor{\eqalign{\phi &= v X(r) e^{2 \pi i \tau /\beta} \cr
                      B_4 &= {2 \pi \over e_R \beta} (P(r) -1) \cr }}
and the  monopole field of the unbroken $U(1)_Q$,
\eqn\mono{A_3 = g_Q (1- \cos \theta),}
with $g_Q$ the magnetic charge. The Dirac quantization condition
implies that $g_Q e_Q = n_M /2$ with $n_M$ integer (we assume that
$e_Q$ is normalized so that the charges $q$, $\hat q$ are integers).
The
non-vanishing
components of the connection $\Gamma_\mu$ are
\eqn\gamcon{\eqalign{\Gamma_1 &=0 \cr
                     \Gamma_2 &= \hf A(r) \gamma^{\hone}
\gamma^{\htwo} \cr
		     \Gamma_3 &= \hf (\sin \theta A(r) \gamma^{\hone}
		                 \gamma^{\hthree} + \cos \theta
\gamma^{\htwo}
				 \gamma^{\hthree} )\cr
		     \Gamma_4 &= \hf A' A \gamma^{\hone}
\gamma^{\hfour} .\cr }}

Corresponding to the $S^2 \times $\real$^2$ structure of the
Reissner-Nordstr\"om black hole manifold, we  would like to decompose
the Dirac
operator $\gamma^\mu D_\mu$ into an angular part depending
on coordinates $(\theta, \phi)$ and a remaining part  depending on
coordinates
$(r,\tau)$. To do this we first write
\eqn\changevar{ \psi(x^\mu) = {W_-(x^\mu) \over \sqrt{A r^2 \sin
\theta}} .}
We then find that
\eqn\spliteq{\gamma^\mu D_\mu \psi = {1 \over \sqrt{A r^2 \sin
\theta}}
          [ A \gamma^{\hone} \partial_r W_- + A^{-1} \gamma^{\hfour}
	  (\partial_\tau + i r_\psi e_R B_\tau ) W_- + {1 \over r} K
W_- ]}
with $K$ given by
\eqn\kdef{ K = ( \gamma^{\htwo} \partial_2 +
{\gamma^{\hthree} \over \sin \theta}
                  (\partial_3 + i Q e_Q A_3)). }
In what follows we take the $Q$ charge of $\psi (\chi)$ to be $-1
(+1)$.

The operator $K$ is essentially the Dirac operator for a charged
fermion
interacting with a $U(1)$ magnetic monopole field on the two-sphere
and
according to the index theorem has $n_M$ normalizable  chiral zero
modes. For
$n_M=1$ the zero mode is given explicitly by
\eqn\ezmode{ W^0_-(\theta ,\phi) = \sqrt{{\sin \theta \over
\tan \theta /2}} \xi_-}
with $\xi_-$ a constant left-handed spinor obeying
$i \gamma^{\htwo} \gamma^{\hthree} \xi_- =-\xi_-$.

We can also decompose $\bar\chi^T$ as
\eqn\xchangevar{ [\bar\chi^T(x^\mu) ]^* = {W_+(x^\mu) \over
\sqrt{A r^2 \sin \theta}} .}
Remembering that $\chi$ has a $U(1)_Q$ charge which is the negative
of
$\psi$ we again find a zero mode of the angular operator of the form
\ezmode\ but with $\xi_-$ replaced by $\xi_+$, a right handed spinor
which obeys $i \gamma_{\htwo} \gamma_{\hthree} \xi_+ = \xi_+$.
Similarly, if we are solving the $(\chi,\bar\psi)$ system, we find
that
$\chi \propto
\xi'_+$, ${\bar\psi}^T \propto \xi'_-$ where $i\gamma^{\htwo}
\gamma^{\hthree}
\xi'_\pm = \pm \xi'_\pm$, but $\xi'_\pm$ are required to have the
opposite
four-dimensional
chiralities to $\xi_\pm$.

We now look for solutions of \euczm\ of the form
\eqn\ansatz{
\eqalign{
\psi &= {W^0_-\over \sqrt{ Ar^2 \sin\theta} }
 f_-(r, \tau), \qquad (\bar\chi^T)^* =
{W^0_+\over \sqrt{ Ar^2 \sin\theta} }
 g_+(r, \tau) \cr
\chi &= {W'^0_+ \over \sqrt{Ar^2\sin\theta}} f_+ (r,\tau), \qquad
(\bar\psi^T)^* = {W'^0_- \over \sqrt{Ar^2\sin\theta}} g_-(r,\tau)
\cr}}
Now, in our basis, we may choose
$C\xi_\pm = \gamma^{\hfour} \xi_\mp$ (in general there will be
some arbitrary but constant phase), and the equality
\eqn\crux{
\gamma^{\hone} \xi_\pm = \mp i \gamma^{\hfour} \gamma^5 \xi_\pm }
implies that inputting these Ans\"atze into \euczm\ yields:
\eqn\fgeqn{
\eqalign{ Af_-' - {2\pi r_\psi \over A\beta} (P-1) f_-
+ {i\over A} {\dot f_-}
+ \lambda v X e^{-2\pi i \tau /\beta}g^*_+ &=0 \cr
Ag_+' - {2\pi r_\chi \over A\beta} (P-1) g_+ + {i\over A} {\dot g_+}
+\lambda v X e^{-2\pi i\tau/\beta} f^*_- &=0 \cr }
}
with the corresponding equation for $f_+$ and $g_-$ having a
minus sign in front of the first term.
Thus, using the $\hat r$ coordinate from \rhat, and noting
that over the range of interest $A\sim 2\pi \hat r/\beta$ and
\xpform\ applies, we see
\eqn\fgpm{
\eqalign{
\mp f'_\pm - {r_\psi \over \hat r} (P_0-1) f_\pm + {i\over \hat r}
{\partial f_\pm \over \partial ({2\pi \tau \over \beta})} +
\lambda v X_0 e^{-2\pi i\tau/\beta} g^*_\mp &=0\cr
\pm g'_\pm - {r_\chi \over \hat r} (P_0-1) g_\pm + {i\over \hat r}
{\partial g_\pm \over \partial ({2\pi\tau\over\beta})} +
\lambda v X_0 e^{-2\pi i\tau /\beta} f^*_\mp &=0 \cr
}}
Clearly this requires $f_- = g_+=0$ for a non-singular solution, and
for
$f_+$, $g_-$ to take the form
\eqn\fg{
\eqalign{
f_+ &=  \sqrt{\hat r} e^{i\alpha} x_+(\hat r) \exp \left \{ \int [
\half
{\textstyle {2\pi (P-1)\over A\beta} }
-\lambda v X ] d \hat r \right \} e^{-\pi i \tau/\beta}\cr
g_- &= \sqrt{\hat r} e^{-i\alpha} x_-(\hat r) \exp \left \{ \int [
\half
{\textstyle {2\pi (P-1)\over A\beta} }
-\lambda v X ] d \hat r \right \} e^{-\pi i \tau/\beta}\cr } }
where $x_\pm$ are, for small $\hat r$, the same functions as appear
in \awkward. Note that these solutions are anti-periodic in the
angular coordinate $ \tau$ as compared to  the solutions \czmode\
which are periodic in $\theta$. This is due to the fact that we have
used
an orthonormal cylindrical basis which is naturally adapted
to the euclidean Schwarzschild solution rather than the cartesian
basis used
in \czmode\ and \jr. For a detailed discussion of the relation
between
these bases see \bw.

Clearly for an anti-vortex/anti-monopole background this solution
will be unchanged, whereas for a vortex/anti-monopole or
anti-vortex/monopole
background the situation will be reversed; it will be $\psi,\bar\chi$
that
has the zero-mode and $\bar\psi,\chi$ which does not.

Notice that the presence of the $\sqrt{A}$ term in the denominator of
\ansatz\ implies that $\chi, \bar \psi$ are of order unity as $\hat
r\to 0$,
i.e., at the horizon. Thus $(\bar\psi,\chi)$ can be
regarded as hair on the black hole (a non-singular non-trivial field
configuration supported by an event horizon) albeit in its euclidean
section.

\newsec{Fermion Hair}

Having established the existence of vortex solutions in a
magnetic Reissner-Nordstr\"om
background and the presence of fermion zero modes when fermions are
coupled as in the model of \witten\ we now wish to discuss
some physical consequences of these zero modes.

The basic formalism for carrying out calculations of the effects of
vortices on black holes carrying discrete charges has been discussed
in some depth in \growhair\ so we shall be rather brief here. The
Euclidean partition function
\eqn\eucpart{Z(\beta) \equiv e^{-\beta F} = \int e^{-S_E}}
for a black hole with both discrete charge and magnetic charge
is to be evaluated by inserting  a projection operator onto states
of definite discrete charge and restricting the path integral
to configurations of definite magnetic flux on the
$S^2$ at radial infinity. The Euclidean path integral
is then saturated by solutions with topology \real$^2 \times S^2$
which
in addition satisfy the constraints imposed by the projection
operator.
In addition the bosonic fields in the path-integral are required to
be periodic in Euclidean time with period $\beta$ while fermion
fields
are required to be anti-periodic.

In our model this corresponds to saturating the functional integral
(ignoring backreaction) with the euclidean  magnetic
Reissner-Nordstr\"om
black hole and a vortex of the broken $U(1)_R$
of vorticity $k$ with $k$ integer. The dominant effects come from
$k=-1, 0, +1$ and the effects of discrete charge appear only in the
sectors with $k \ne 0$.

The first immediate consequence of the zero modes we have found
is the vanishing of the $k \ne 0$ contributions to the partition
function  and to all other correlation functions which do
not involve the requisite number of fermion fields. It follows
in this model that the temperature of the black hole is not
modified by the presence of discrete charge, nor does the  screened
electric
field acquire a exponentially small probability outside the horizon
of the black hole - it is completely extinguished.

The only operators which will acquire non-zero expectation values
in the vorticity $k=\pm 1$ sectors will be those with the correct
number of fermion operators to soak up the zero modes.
With a single $(\psi, \chi)$ multiplet and a single $(\hat \psi, \hat
\chi)$
multiplet the simplest anomaly free theory (as mentioned in sec. 2)
has
$\hat q=q$ and $\hat r = r+1$. If we choose $r=1/N$ then
the $U(1)_R$ symmetry is broken down to \zed$_N$ and we can
in the usual way assign a \zed$_N$ charge to black holes.
In this theory we find that the vortex
sector gives rise to zero modes for $(\bar \chi, \psi)$ and
$(\bar{ \hat {\psi}\,}, \hat \chi)$ leading to an
expectation value for the operator ${\cal O} = \psi^T C \hat \chi
\bar{\hat{\psi}\,}
C \bar \chi$
while the anti-vortex sector
gives an expectation value to the hermitian conjugate operator. Note
that these operators are neutral under the unbroken $U(1)_Q$ as would
be expected for a non-anomalous symmetry.

The situation described above is reminiscent of the the instanton
calculation
of the $\theta$-dependence of correlation functions in massless QCD
with the discrete charge of the black hole playing the role of the
the
$\theta$-parameter of QCD. In massless QCD the proper interpretation
is that the $\theta$-vacua are no longer physically distinguishable
but rather
label the globally degenerate vacua resulting from spontaneous
breaking
of the $U(1)_A$ axial symmetry by the QCD vacuum. The key to this
identification is the chiral anomaly
\eqn\chiranom{\partial_\mu j^{\mu 5} = {g^2 \over 16 \pi^2}
 {\rm Tr} F \tilde F}
and the resulting chiral Ward identities \ref\colebook{ for a
pedagogical
discussion see S. Coleman, The Uses of Instantons, in Aspects of
Symmetry,
Cambridge University Press, Cambridge 1985.}.

We should inquire whether a similar mechanism is operative here.
The answer in the models we have examined is yes. To see how this
arises in the above example first note that the theory also possesses
a global symmetry $U(1)_X$ which is orthogonal to both $U(1)_R$ and
$U(1)_Q$ with $\psi$ and $\hat \chi$ having $X=1$ and $\chi$ and
$\hat \psi$
having $X=-1$. This global symmetry is however anomalous as can be
seen by noting that e.g.~${\rm Tr} X Q R = -2q$. It is also clear
that the operator ${\cal O}$ which acquires
a non-zero expectation value transforms non-trivially under $X$.
We thus find that the anomalous $U(1)_X$ symmetry is spontaneously
broken
in the black hole background and hence any putative discrete charge
can be
absorbed by performing
a $U(1)_X$ transformation.

We thus come to a rather surprising conclusion in this class of
theories.
In the background of a magnetically charged black hole there is no
physical effect of discrete charge, rather the space outside the
horizon
of the black hole is filled with a fermion condensate which violates
a global, anomalous symmetry. In contrast, if we were to consider
an electrically charged black hole we would find no fermion zero
modes, there would be no fermion condensate, and one could measure
the effects of discrete charge as usual. It is striking that electric
and magnetic black holes which are so similar classically can have
such distinct quantum structure.

In semi-realistic models such as the $O(10)$ vortex model of \witten\
we would generically find a fermion condensate outside the black hole
which violates baryon plus lepton number. This condensate would
presumably lead to baryon number violating scattering processes off
the
black hole which are in some ways analogous to the Callan-Rubakov
effect
except that here they would be driven by the anomaly in the baryon
number current rather than by superheavy $X$-boson exchange in the
core of the monopole.

It is worth emphasizing two points regarding the fermion hair we have
discovered. The first is that it is secondary hair in the language
of \growhair. That is, it does not enlarge the space of quantum
states
of the black hole but is rather determined by the quantum numbers
(in this case magnetic charge) carried by the black hole.
The second point
is that this hair, unlike that discussed in \growhair, is not
dependent on the existence of discrete gauge charge for its
existence. In fact, as argued above, the condensate removes the
final wisps of discrete gauge hair and replaces them with a rich mane
of
fermionic hair.

\bigskip\centerline{\bf Acknowledgements}\nobreak

This work was supported
in part by NSF grants PHY90-00386 and PHY 89-18388. R.G. was also
supported
by the McCormick Fellowship fund at the Enrico Fermi Institute; J.H.
acknowledges the support of NSF PYI Grant PHY-9157463.

\listrefs
\end